\documentstyle[12pt]{article}
\def\papertitlepage{ \thispagestyle{empty}}
\def\eqbegin         {  \begin{eqnarray}  }
\def\eqend           {  \end{eqnarray}  }
\def\sectionnumbering {\setcounter{equation}{0}\renewcommand{\theequation}{\arabic{section}.\arabic{equation}}}
\def\lamda        { \lambda }
\def\iPhi         { {\mit \Phi}}  
  
\def\iDelta       { {\mit \Delta}}
\def\iLamda       { {\mit \Lambda}}
\def\del          { \partial }

\begin{document}
\papertitlepage
\vspace{30mm}

\begin{center}

{\large \bf Chiral Vertices,Fusion Rules and  Vacua of  Fractional Quantum Hall Systems} 
\vspace{60mm}

{\bf Kazusumi Ino}
\end{center}

\begin{center}
{\sf Institute for Solid State Physics, University of Tokyo,

 Roppongi
 7-22-1,Minatoku,Tokyo,106, Japan} \\
{\rm Tel: +81 3 3478 6811;Fax:+81 3 3478 4903}
\end{center}

\centerline{\sf ino@momo.issp.u-tokyo.ac.jp}

\vspace{10mm}

\begin{abstract}
Vacua  of two-dimensional incompressible systems,such as FQH systems,  are
 characterized by rational conformal field theories.  
 We develop a method to express 
the wavefunctions of FQH-like systems in terms of chiral vertices.
We formulate quantum mechanics on those expressions, which reveals the  
simple structure underlying the conformal field theory description
of the quantum hall states.  Also we argue the recent conjecture of 
Nayak and Wilczek   
on the spinor statistics of 2n quasihole state in paired quantum hall states.

\end{abstract} 



\newpage
\newcounter{1}[page]
\section{Introduction}
\sectionnumbering

In the universe, various kinds of low dimensional structures exist.  For
 more than a decade physicists have been paying much effort  to explore those
 low dimensional structures.  The regions where low dimensional structures 
emerge  widely spread in many branches of  physics. 
  High temperature superconductivity and quatum hall effect 
in condensed matter systems, string theory which unifies 
 space-time and particle, and various solvable
 statistical models and their critical behavior  and so on.
 In all cases, {\it two} dimensional nature of system 
plays a prominent role.  The common  feature is that two dimensional structure
 seems to have a deep relationship with the phases of the system.  

Among them, Fractional Quantum Hall Effect (FQHE)\cite{Tsui} is remarkable
 from experimental verifiability.  In the 
experiments for this system  one can control arbitrarily the parameters 
which govern the vacuum of the system, that is the strength of a background 
uniform magnetic field and the electron's density. 
The vacua are parametrized by the ratio of electron's density and 
the strength of the magnetic field, which is called 'filling fraction'.

Fractional quantum hall effect was first discovered by Tsui, Stormer and 
Gossard in 1982\cite{Tsui} in the two-dimensional system of electron 
in a uniform magnetic field with  odd denominator filling fractions. 
It was first phenomenologically understood by the 
 first-quantized approach initiated by Laughlin \cite{Lagh}. 
The remarkable ingredient is
 {\it fractional statistics} obeyed by 
the  quasiparticles sustained by the vacua. 
Thus fractional quatum hall effect gives  us the evidence  that, in 2+1 dimension,
 there are very rich structure of vacua. These approach was extended to
so called hierarchical construction to include the general odd denominator 
filling fractions\cite{Hal} \cite{Hal2}.

Some years later their dynamics are formulated as a 2+1 dimensional system
 of second-quantized particle coupled to U(1) Chern-Simons gauge field. 
Chern-Simons gauge field adds a flux to the particle to which the gauge field couples.
 The most transparent formulation is through Landau-Ginzburg type theory
\cite{SCZ}.
In LG formulation,  the quasiparticle appears as a classical configuration  of 
 charged boson and U(1) Chern-Simons  field.  

As they are formulated with Chern-Simons gauge field,  the relation between 
 FQHE and  conformal field theories  naturally emerges.  In  fact 
in \cite{MorRe}, it was shown that various FQH wavefunctions could be 
interpreted as  certain conformal blocks of proper RCFT. 
They  proposed that the effective theory for the bulk  is
 the corresponding  Chern-Simons theory.
To prove this proposal, it should be verified 
 quasiparticle's statistics (defined by Berry's  phase)
 are given by the braiding property obtained from  
 Chern-Simons theory.
Although this is proved for abelian cases,
 many indications show the validity of  this proposal.
For examples, \cite{BW} gives an argument for the system obtained from 
$SU(N)_k$ model and, more recently \cite{NW} gives
 a general argument in favor of this proposal. 
Assuming that this is true, some FQH states are supposed to have 
a excitation  with so-called {\it nonabelian statistics}.
These states are  some paired quantum hall states \cite{HalRez} \cite{Grei}
 which might be realized in FQH systems with even-denominator
 filling fractions \cite{JE}.  Recently, those paired quantum hall states have
 attracted much attention \cite{Milo}--\cite{GFN}. 
Other examples of nonabelian statistics were given in \cite{BW} based on 
$SU(N)_k$ Wess-Zumino-Witten models.

There are other kind of formulation of FQHE,  
developped in \cite{AC}--\cite{AC2},
which is the  approach to the formulation of FQHE from
 area-preseving diffeomorphism.
This approach is based on  the incompressible liguid nature of FQH system. 
A hierarchical constuction scheme, which fit to 
composite fermion  construction  of \cite{Jai},
 arises from  the existence of particular class of 1+1 dimensional 
field theory($W_{1+\infty}$ minimal model (this is not RCFT )) on the edge.
  Also  excitations with nonabelian statistics are found on these models.

FQHE has been investigated in a system of electron. However, recently,  
the possiblity of FQHE also in a bosonic system   was pointed out \cite{Jain}.
This is natural because, as the coupling to Chern-Simons gauge field
 changes the statistics of particle, the original statistics of the particle does 
not matter.

In this paper we will study the basic aspects of 
general  2+1 dimensional incompressible systems
 described by  rational conformal field theories. 
We will use the canonical quantized formulation
 of RCFT.This is because there is a system  
which lacks a suitable dynamical 2+1 dimensional field theory description. 
We introduce a method to express the wavefunctions of
 FQH-like systems in terms of chiral vertices. We define 
some useful operator on chiral vertices.   
We also argue  the recently conjectured       
spinor statistics of Pfaffian states \cite{NW} by these formulation.

 Our main philosophy is that RCFT (or its operator algebra) characterizes 
the vacua of 2+1 dimensional incompressible system . 
Fusion algebra \cite{Ver} is the most essential 
ingredient for the system characterized by rational conformal field theory.
\footnote{The role of fusion algebra in FQH systems 
 were previously discussed in \cite{MorRe}\cite{WW}. }

The organization of this paper is as follows.
In sect.2 we review generalities  
on rational conformal field theory. 
Sect.3 describes the general feature of 
canonically quantized RCFT as the incompressible system. In particular, 
 we formulate the  extended U(1) current algebra
 description of a background uniform magnetic field, which will 
be useful when one deals with a charged particle system such as FQHE. 
In sect.4  we discuss the property of some operators  on chiral vertices.
Sect.5 gives some examples, which
 involves some known states of FQHE. 
Also we describe how one can deduce  the spinor statistics of multiple 
quasihole Pfaffian state. 

\section{Generalities on  RCFT}
\sectionnumbering

     First we recall some general properties of 
rational conformal field theory(RCFT). 
Let $\Sigma$ be a Riemann surface of genus $g$ with n punctures $z^{i},i = 1\cdots$n.
We denote their moduli by $m^j,j=1,\cdots,3g-3$.
A conformal field theory is said to be rational if its unnormalized 
n-point function $G$
has an expansion 
\begin{equation}G(z^{i},m^{j},\overline{z}^{i},\overline{m}^{j})
=\sum_{I J}h_{\overline{I}J}
\overline{\cal F}_{\overline{I}}(z^{i},m^{j}){\cal F}_{J}(\overline{z}^{i},
\overline{m}^{j})
\end{equation}
 as a sum of products of holomorphic building blocks  ${\cal F}_{I}$ and antiholomorphic ones
$\overline{\cal F}_{\overline{J}}$, and $h_{\overline{I} J}$is a hermitian metric.
${\cal F}_{I}$  span a finite dimensional vector space. 
In general,  ${\cal F}_{I}$ have a nontrivial monodromy and modular behavior.  
These properties are further specified by  some PDE
which arise from the apperance of null  states.  ${\cal F}_{I}$  are in fact  a
  holomorphic sections of  a  vector bundle ${\cal V}_{g,n}$ over the moduli space
 ${\cal M}_{g,n}$ of punctured Riemann surfaces. 
As ${\cal V}_{g,n}$ is finite dimensional, rational 
conformal field theories  have only a finite number of Virasoro representations i.e. 
primary fields. 

In this paper we don't treat the properies associated with the hermitian metric 
$h_{IJ}$ although these properties are  very remarkable . We will mainly  
treat chiral half of RCFT in the remainder of this paper. 

Let us consider some RCFT with $N$ primary fields $\phi_{i},(i=0,\cdots N-1)$ 
corresponding to the irreducible representations $[\phi_{i}]$.
$\phi_{0}$ is always used to denote the representations which contains 
the identity. In this RCFT any analytic blocks are built out from three point 
functions $\langle\phi_{i}\phi_{j}\phi_{k}\rangle$ by so-called
 sewing (or gluing) procedure.  The sewing procedure  can be 
 formally written by Feynman diagrams of ${\phi}^{3}$ theory by relating 
the ${\phi}^{3}$ vertices to $\langle\phi_{i}\phi_{j}\phi_{k}\rangle$ .  
However the correspondece between Feynman diagrams of ${\phi}^{3}$  and the 
correlation functions  on  punctured Riemann surfaces with operators at punctures 
is not one to one. $\phi^{3}$ graphs are redundant. 
This redundancy leads to {\it duality} between different bases of analytic blocks.

To describe duality (and for later convenience),  it is convenient to use the concept 
{\it chiral vertex}(see, for example \cite{Schro}).
 It is the operator  which
 is defined when we see a representation $[\phi_{i}]$ as inducing
 a  operator ${\iPhi}_{jk}^{i}(z): [\phi_{k}] \rightarrow [\phi_{j}] $. 
 This definition is specified by the matrix element between primary fields
:$ \langle j|\iPhi_{jk}^{i}(z)|k \rangle=
t_{jk}^{i}z^{-\iDelta_{t}},\iDelta_{t}=\iDelta_{i}+\iDelta_{k}-\iDelta_{j} $.  
Here $t_{jk}^{i}$   are invariant tensors of the theory. 
Its actions on descendants are deduced from this formula.
This chiral vertex itself represents the sphere with $\phi_{i}$ at $z$,
$\phi_{k}$ ($\phi_{j}$) at the origin (the infinity). 

There are two fundamental operations on chiral vertices. 
One of the fundamental operations is {\it fusing} \cite{MorSei}
\footnote[1]{For simplicity, 
we use the notation which is valid for the case where the fusion rule coefficient 
$N_{ij}^{k}$ (defined in the latter part of this section) are 0 or 1.   Typical 
models ($SU(2)$ WZW models, Ising model etc) satisfy this condition.}:
\eqbegin 
\iPhi_{kp}^{i}(z_1)\iPhi_{pl}^{j}(z_2)=\sum_{q}F_{pq}\left[\begin{array}{cc} i&j\\
 k&l\end{array}
\right]\sum_{Q} \langle Q|\iPhi_{qj}^{i}(z_{12})|j\rangle\iPhi_{kl}^{q,Q}(z_{2}),
\label{eq:F}\eqend
where $z_{12}=z_1-z_2$. This is nothing but the operator product expansion. The other 
 fundamental operation is {\it exchange} or {\it braiding}{\cite{Schro}:
\eqbegin 
\iPhi_{kp}^{i}(z_1)\iPhi_{pl}^{j}(z_2)=\sum_{q}R_{pq}\left[\begin{array}{cc} i&j\\ 
k&l\end{array}
\right]\iPhi_{kq}^{j}(z_{1})\iPhi_{ql}^{i}(z_{2}) 
\label{eq:B}.\eqend
$R$ depends on which region one compare the blocks.   We will take   
$R$ in ${\rm Im}(z_1-z_2)> 0$.
 $R$ is related  to the  monodromy of analytic blocks. In fact\\ 
$
\langle k|\iPhi_{kp}^{i}({\rm e}^{2\pi i}z)\iPhi_{pl}^{j}(0)|l\rangle = $
\eqbegin
\sum_{q,s}
R_{ps}\left[\begin{array}{cc} k&j\\ i&l\end{array}\right]
R_{sq}\left[\begin{array}{cc} j&k\\ i&l\end{array}\right]
\langle k|\iPhi_{kq}^{i}(z)\iPhi_{ql}^{j}(0)|l \rangle .
\eqend
$F$ and $R$ can be casted  as  base change  transformations in the space of 
analytic blocks. This interpretation comes from the fact that chiral vertices 
represent the three-point functions on the sphere and analytic blocks 
are built up from these functions.
  More precisely this transformation are formulated as
 the bundle transformations of the vector bundle over the  moduli space
$\overline{\cal M}$ as first formulated by Friedan and Shenker \cite{FS}. 
In terms of  $F$ and $R$,  
the condition for the equivalence of different descriptions of 
analytic blocks, i.e. duality 
is reduced to the equations among $F$ and $R$ and  conformal weight.
These are polynomial equations  such as pentagon and hexagon relations of
 \cite{MorSei}. 
 From those relations $R$ can be written in terms of $F$ as 
\eqbegin 
R_{pq}\left[\begin{array}{cc} i&l\\k&j\end{array}\right]  
=e^{i\pi (\iDelta_{k}+\iDelta_{j}-\iDelta_{p}-\iDelta_{q})}F_{pq}
\left[\begin{array}{cc}
 i&j\\ k&l\end{array}
\right] .
\label{eq:BF}
\eqend
Thus the exchange property is determined from the fusing property and conformal 
weights of primary fields.

The concept of chiral vertex  will be  essential for nonabelian excitations
discussed in  later sections. 

Finally let us recall  fusion rules.  Let ${\cal V}_{0,ijk}$ be
 the vector bundle corresponding 
to the three puctured sphere with $\phi_{i}, \phi_{j}, \phi_{k}$ at the punctures. 
We put $N_{ijk}$ to be dim${\cal V}_{0,ijk}$ .  The indices are raised by $N_{ij0}$.  
Actually $N_{ij0}$ is conjugate matrix $C$ and $N_{ij}^{k}$ is the dimension
 of the space 
of $t_{ij}^{k}$. Then fusion rules are defined as 
\begin{equation}\phi_{i} \times \phi_{j} = \sum_{k} N_{ij}^{k} \phi_{k} 
\label{eq:fusion}\end{equation}
(\ref{eq:fusion}) shows how many ways there are to fuse $[\phi_{i}]$
 and $[\phi_j]$ to 
$[\phi_{k}]$.  From duality we have following equation, 
\eqbegin \sum_{k} N_{ij}^{k} N_{klm}=\sum_{k} N_{il}^{k} N_{kjm}
\label{eq:duality}. \eqend 
This relation (\ref{eq:duality}) implies the associativity of the fusion rules.
Fusion rules therefore forms a representation of chiral ring. 
A highly nontrivial fact is this representation is determined 
by the modular property of Virasoro characters \cite{Ver}.
 Famous Verlinde 
formula  states that fusion rules have  a deep relation with
 the modular behavior of Virasoro 
characters.

The degeneracy of analytic  blocks on a general
 Riemann surface  can be computed from the fusion rules.  By  $\phi^3$-diagram
 representation of the given analytic block we can built up any 
analytic blocks
 on Riemann surfaces. From the definition of fusion rules and duality,
 this description gives a consistent number for degeneracy.
By this method,  we have general formulas  for degeneracy as follows:

\begin{equation} \begin{array}{cccc}
       {\rm  amplitude  } & {\rm \  Degeneracy  \    on }    S^2   &  {\rm on}   T^2 &
\cdots    \\   
 \langle \phi_i\phi_j\phi_k \rangle  & N_{ijk}&N_{im}^{l}N_{ln}^{m}N_{jk}^{n}&\cdots\\
 \langle \phi_i\phi_j\phi_k\phi_l \rangle & N_{ijm}N_{kl}^{m}& N_{ij}^{p}N_{pq}^{r}
N_{rs}^{q}N_{kl}^{s}&\cdots \\
\cdots&\cdots&\cdots&\cdots
\end{array}.
\end{equation}

For more details of RCFT, see \cite{MorSei}.

\section{2+1 Dimensional System and RCFT}
\sectionnumbering
\subsection{Generalities}

We will  consider how a given RCFT looks like from  2+1 dimensional point
 of view 
in this section.

The most famous connection between RCFT and 2+1 dimensional physics is  
 through Chern-Simons theory \cite{Witt}. It is known that 
chiral algebra of RCFT can be obtained from quantizing
 Chern-Simons theory on the disk and conformal blocks 
 can be obtained from Chern-Simons theory with Wilson lines \cite{MorSei2}. 
Chern-Simons theory is useful to argue qualitive properties of RCFT. 
However we'd rather   give a basic characterization of RCFT as 
incompressible system  in this section.

As in the last section,  let $[\phi_{i}]$ be the  Virasoro representations of
 the chiral half of some RCFT. These fields may represent excitations of 
2+1 dimensional system.  First of all a Virasoro representation
 [$\phi_i$] should correspond to  a 2+1 dimensinonal {\it single}
 particle excitation.
When we apply some infinitesimal conformal reparametrization on 2 dimensional
 space, primary field $\phi_i$ transforms to  a linear combination of 
 descendant fields.
Since this is merely a effect of reparametrization of space,  they should be
 physically the same.  This is the first reason for the above statement.  
Also  it is seen in the previous section that $F$ and $R$ is 
defined for Virasoro representations.   
Since those operations give us  the statistical property in 2+1 dimension
(see sect 3.2),  again a Virasoro representation
 [$\phi_i$] should correspond to  a 2+1 dimensinonal single
 particle excitation.
Thus it is natural to see  a  Virasoro representation $\phi_i$ 
as  $\it quasiparticle$ in  2+1 dimension. 
We will consider these $\phi_{i}$ as a quasiparticle excitation  of
some 2+1 dimensional system.

We next consider  what kind of   'hamiltonian' we should
 take for this system. 
We  consider a  second-quantized (in a sense) heuristic hamiltonian   
which contains 
the essential physical 
content  the given RCFT has as a 2+1 dimensional object.
Among various CFT operator the most simple choice is : 
\eqbegin 
H=L_{0} 
\label{eq:H}.
\eqend
Although this seems to be too simple, we will see in sect.3.6  
that we must use the projection for conformal weight 
 to form a suitable 2+1 dimesional  quantum mechanical hamiltonian $\cal H$. 
In that description  (\ref{eq:H}) sees
 the essence of qualitive nature of the system described by 
a given RCFT. 
({\ref{eq:H}}) is also the hamiltonian of 1+1 dimensional physics on the edge
 and measures the energy of 1+1 dimensional excitations.

However, as we regard a conformal family corresponding to a single
 quasiparticle in the bulk, 
we must project the state to some representative state before
 we apply $H$ as in the way of gauge fixing. 
To this end  we may simply take the projection operator $P_L$ 
 to the lowest level state of conformal family: 
\eqbegin 
M_0=P_L H P_L 
\label{eq:mass}.
\eqend
By taking  this projection, we can reduce descendant fields for this 
operator.   Also we can see conformal blocks as representatives
 of physically equivalent analytic blocks.
So we will  treat $M_0$ as a heuristic ' hamiltonian ' or
 a {\it mass} of excitations.  This is somewhat 
similar to the situation in string theory. 
 By taking $M_0$, the vacuum supports a gap since this is RCFT,
it means this system is {\it incompressible}.

Quasiparticles  $\phi_{i}$ 's short-distance behaviors or interactions 
 are obtained  from the operator product expansion.  However in the
 long-wavelength limit 
one may ignore the detail of short-distance physics, and an information
 which 
neglects such details is essential. Such a information 
is provided by the fusion rules (\ref{eq:fusion})
\begin{equation}\phi_{i} \times \phi_{j} = \sum_{k} N_{ij}^{k} \phi_{k} 
.\label{eq:fusion2}\end{equation} 
Although there is the case when different CFT have a same fusion rules,   
fusion rules  are at least  decisive for whether the system have abelian 
statistics or nonabelian statics and what kinds of bound state exist in the system. 
It is  known that 
fusion rules lead to   a rather hard restriction on the central charge and 
conformal weight.

Let us see how (\ref{eq:fusion2}) arises.
The meaning of (\ref{eq:fusion2}) is from Deligne-Mumford 
stable compactification of moduli space $\overline {{\cal M}_{g,n}}$.  
The process in which two points $z_1$ and 
$z_2$ come close to each other can be described as the process in which a sphere, 
that contains $z_1$ and $z_2$ at fixed distance, 
pinches off the surface by forming  a neck 
of length log$|z_1-z_2|$. The end of this process is the surface with $z_1=z_2$ and a sphere 
with two  vertex operators $\phi_i (z_1),\phi_j (z_2)$ plus one extra marked point where 
another vertex operator $\phi_k$  is inserted. This is why $N_{ij}^k$ must
 appear on the
 righthand side of (\ref{eq:fusion2}). This process also clearly shows how
 chiral vertices $\iPhi_{ij}^{k}$ appears.

For a  Virasoro representation corresponding  to a quasiparticle of 2+1 
dimensional system, a conformal block
 $\langle \phi_{i_1}(z_1)  \cdots \phi_{i_n}(z_n)\rangle $ is  considered as 
the amplitude for quasiparticles to be exist at  $z_1 \cdots z_n$. This means
$\langle \phi_{i_1}(z_1)  \cdots \phi_{i_n}(z_n)\rangle $ may be 
 the wavefunction for some many-body system.
  A   meaning   of  chiral vertex 
 $\iPhi_{ij}^k$ in this context is that it is a  expression   that 
represents concisely the situation when $\phi_k$ is on some place
 within a region 
conformally isomorphic to  annulus or tube and its boundary conditions  or 
asymptotic Virasoro representations  for  path-integration are
 $\phi_i$ and $\phi_j$. 
On each region, we consider the 
local Hilbert space for the given RCFT over  $SL(2,C)$ invariant vacuum.
 Each region is mapped to the 
Riemann sphere with three marked points by a local conformal map 
(in the limit i.e. boundaries mapped to points. In this limit we are in 
$\overline{{\cal M}}$ again.).

From this viewpoint on chiral vertices, let us  consider
 the amplitude for  $n$ excitations $\phi_{i_1},\cdots \phi_{i_n}$
 on a closed  Riemann surface $\Sigma$  and introduce the  {\it canonical 
 expression } of $\langle \phi_{i_1}(z_1)  \cdots \phi_{i_n}(z_n)\rangle $
   in terms of chiral vertices. 
First we make a simply connected
 region on $\Sigma$ which only contains the point at which the 
 excitation $\phi_{i_n}$ is,  and no other excitations. 
Second, we make a region isomorphic to annulus 
which only contains $\phi_{i_{n-1}}$. We continue this procedure, 
to end up with 
the simply connected region which contains $\phi_{i_1}$. 
   When $\Sigma$ has  
 genus $\geq 1$, we must insert $2g$  region  $C_1,\cdots, C_{g}$, 
$D_1,\cdots ,D_{g}$ with three boundaries, which  don't contain 
 any excitations. 
This procedure 
defines one direction of  {\it flow} on $\Sigma$. This flow must not be  
stopped up except at punctures  or real boundaries (in case  they exist).
$C_j$ arises when the flow on $\Sigma$ splits into two, and $D_j$ arises 
when two flows  joint into one.  By this procedure 
$\Sigma$ is divided into n regions $A_1,\cdots,A_n$ and
  $2g$  region  $C_1,\cdots, C_{g}$, $D_1,\cdots , D_{g}$ . 
Now we can express the given amplitude  through chiral vertices. 
The amplitude  $\langle \phi_{i_1}\cdots \phi_{i_n}\rangle $ corresponds
 to the expression  of chiral vertices 
\eqbegin 
\iPhi_{1 a_1}^{i_1}\cdots
\overline{\iLamda}_{d_{3j-2} d_{3j}}^{d_{3j-1}}\cdots 
\iPhi_{a_{2l-2} a_{2l-1}}^{i_{i_l}} \cdots 
\iLamda_{c_{3k-2} c_{3k}}^{c_{3k-1}}\cdots
\iPhi_{a_{2n-2} 1}^{i_{n}},
\label{eq:Gmono}
\eqend 
where $\iLamda_{c_{3k-2} c_{3k}}^{c_{3k-1}},k=1,\cdots,g $ are from $C_k$ and 
 $\overline{\iLamda}_{d_{3j-2} d_{3j}}^{d_{3j-1}}$ are from $D_j$. 
All $a_i$ and $c_m$ and $d_m$ should be 
sewed to some $a_i$ or $c_k$ or $d_k$. 
$a_{2l}$ are   sewed to $a_{2l-1}$ or $d_{3j}$ or $d_{3j-1}$,   
 and  $a_{2l+1}$ are sewed to $a_{2l+2}$ or $c_{3j}$ or $c_{3j-1}$.
  $c_{3k-2}$ and $c_{3k-1}$ are sewed to $a_{2i+1}$ or $d_{3j}$ or $d_{3j-1}$,
 and  $c_{3k}$ are sewed to $a_{2i}$ or $d_{3j-2}$.  
By the above division of $\Sigma$ we know which of $a_i$, $c_k$ and $d_l$  
should be the same and sewed.
  This amplitude has a degeneracy which is  computed 
from the fusion rule coefficients of the given RCFT. 
 On a sphere they can be simply written as 
\eqbegin 
\iPhi_{1 a_1}^{i_1}\iPhi_{a_{1}a_{2}}^{i_{2}} \cdots
\iPhi_{a_{n-2}a_{n-1}}^{i_{n-1}}
\iPhi_{a_{n} 1}^{i_{n}}.   
\eqend
 The different ordering of $\phi_i$ gives other expression. 
However the property of duality reviewed 
in the previous section ensures all of those expressions lead to the same
 amplitude.             

Also when  the system is on a Riemann surface with some boundary, we
 can do the 
same procedure described above with the orientation of the boundaries. 
This again gives us the canonical expression of the amplitude. 
 Let us  consider a Riemann surface with some boundaries,say a disk $D$.
 Then  we can do the procedure such that 
the  region $A_1$  has a real boundary $\partial D$. 
Now $A_1$ is not simply connected region 
but becomes an annulus-like region. The amplitude now corresponds to 
\eqbegin 
\iPhi_{b a_1}^{i_1} \iPhi_{a_1 a_2}^{i_2} \cdots\iPhi_{a_{n-2}a_{n-1}}^{i_{n-1}}
\iPhi_{a_{n-1} 1}^{i_{n}},
\eqend 
where $b$ represents the boundary state.
In the case of an annulus,  
the expression becomes like  
\eqbegin 
\iPhi_{b_1 a_1}^{i_1} \iPhi_{b_2 a_2}^{i_2} \cdots
\iLamda_{c_1 c_3}^{c_2}\cdots\iPhi_{a_{2n-5}a_{2n-4}}^{i_{n-1}}
\iPhi_{a_{2n-3} 1}^{i_{n}},
\eqend 
where $b_1,b_2$ represent  the two boundary states of the annulus. 
These two states 
are constrained by each other through the fusion rules.  
Therefore the fusion rules of the theory gives a  correlation of
 the two edge states.
When there are 
$m$ boundaries, by inserting $\iLamda_{c_{3k-2} c_{3k}}^{c_{3k-1}},k=1,
\cdots,m-1 $, 
we get the same kind of the expression of 
the amplitude in terms of  chiral vertices
\footnote[1]{This   
 canonical expression have a   similarity with  
'temporal gauge' introduced  in 2D gravity \cite{2dTemp}. }.
The multiple boundary states are again correlated through the fusion rules.

In this expression,
the creation of quasiparticle $\phi_{i}$ at $z$ on some state is achieved by 
$\iPhi_{jk}^{i}(z)$ where $j$ and $k$ should be suitably chosen (depends 
on the state on which $\phi_i$ is created.). 

Now let us define  'mass' operators
 for chiral vertices,  which will be shown to have a direct relation to
 2+1 dimensional 
many-body system interpretation in sect.4.  By (\ref{eq:mass}),  
we define $M$ and ${\overline M}$ as follows: 
\eqbegin 
M\iPhi_{jk}^{i}=(\iDelta_i-\iDelta_j+\iDelta_k)\iPhi_{jk}^{i} ,   \\ 
M\iLamda_{jk}^{i}=(\iDelta_i-\iDelta_j+\iDelta_k)\iLamda_{jk}^{i} ,  \\  
M\overline{\iLamda}_{jk}^{i}=(-\iDelta_i-\iDelta_j+
\iDelta_k)\overline{\iLamda}_{jk}^{i} , \\   
\overline{M}\iPhi_{jk}^{i}=(\iDelta_i+\iDelta_j-\iDelta_k)\iPhi_{jk}^{i}
 ,  \\  
\overline{M}\iLamda_{jk}^{i}=(\iDelta_i+\iDelta_j-\iDelta_k)\iLamda_{jk}^{i}, 
 \\  
\overline{M}  \overline{\iLamda}_{jk}^{i}=(-\iDelta_i+\iDelta_j-
\iDelta_k)\overline{\iLamda}_{jk}^{i} ,    
\label{eq:Mass}
\eqend  
if $N_{jk}^{i}\ne 0$, and $0$ othewise,  where $\iDelta_l$ are the eigenvalue of 
$M_0$ defined in (\ref{eq:mass}), 
\eqbegin 
M_0 \phi_l=\iDelta_l\phi_l. 
\label{eq:m_0}
\eqend  
We also adopt (\ref{eq:m_0}) to the state at the boundaries. This means that
 we only 
treat the state at the boundaries like an asymptotic state of
 quasiparticle in the
 bulk when we apply $M$ and ${\overline M}$. We assume $M$ and $\overline M$
 satisfy  Leibniz rule when we apply them  to a monomial
 $\iPhi$, $\iLamda$ and $\overline{\iLamda}$.  
It implies, for example,  

$M\left[\iPhi_{b_1 a_1}^{i_1}\cdots
\overline{\iLamda}_{d_{3j-2} d_{3j}}^{d_{3j-1}}\cdots  
\iLamda_{c_{3k-2} c_{3k}}^{c_{3k-1}}\cdots
\iPhi_{a_{2l-2} a_{2l-1}}^{i_{i_l}} \cdots 
\iPhi_{a_{2n-2} 1}^{i_{n}} \right] $
\eqbegin 
 =(\sum_{k} \iDelta_{i_k}-\sum_{m}\iDelta_{b_m}) 
\left[\iPhi_{b_1 a_1}^{i_1}\cdots
\overline{\iLamda}_{d_{3j-2} d_{3j}}^{d_{3j-1}}\cdots 
\iLamda_{c_{3k-2} c_{3k}}^{c_{3k-1}}\cdots
\iPhi_{a_{2n-2} 1}^{i_{n}}\right],
\label{eq:Cal1}
\eqend 

$\overline{M}\left[\iPhi_{b_1 a_1}^{i_1}\cdots
\overline{\iLamda}_{d_{3j-2} d_{3j}}^{d_{3j-1}}\cdots 
\iPhi_{a_{2l-2} a_{2l-1}}^{i_{i_l}} \cdots 
\iLamda_{c_{3k-2} c_{3k}}^{c_{3k-1}}\cdots
\iPhi_{a_{2n-2} 1}^{i_{n}}\right] $
\eqbegin 
 =(\sum_{k} \iDelta_{i_k}+\sum_{m}\iDelta_{b_m}) 
\left[\iPhi_{b_1 a_1}^{i_1}\cdots
\overline{\iLamda}_{d_{3j-2} d_{3j}}^{d_{3j-1}}\cdots  
\iLamda_{c_{3k-2} c_{3k}}^{c_{3k-1}}\cdots
\iPhi_{a_{2n-2} 1}^{i_{n}}\right],
\label{eq:Cal2}
\eqend 
where all $a,c,d$ should be sewed to each other plausibly explained as
 above,and 
$b_i$ are the states on the boundaries. From these formula,  
$M$ and $\overline{M}$ depend only 
on $\phi_{i_k}$ and the states on the boundaries $\phi_{b_m}$. 
These  operators act invariantly under duality.
Expressions  like  (\ref{eq:Gmono}) are  also  eigenstates  for $M$ and
 its eigenvalue 
is  the sum of conformal weights   of its excitation.
We will see this operator has a  relation with 2+1 quantum mechanical
  interpretation in sect.4.

The discussion in this section can  also be extended to antichiral half of 
RCFT.

\subsection{Statistics of Quasiparticles}

To argue the definition of statistics of quasiparticles, we must first 
recall the fact that the fundamental object in CFT 
is not a single operator $\phi_i$.
The fundametal object is the sphere with 3 inserted operator,
that is represented by chiral vertex $\iPhi_{jk}^{i}$ . 
So statistics should be considered on $\iPhi_{jk}^{i}$, not on
 a single Virasoro representation $[\phi]$.  
For example  we will see the statistics in Pfaffian state 
 can be naturally understood as a property of 
 chiral vertex  $\iPhi_{jk}^{i}$,not on $[\phi_i]$. 

Statistics of quasiparticle is defined by Berry's phase. 
Whether this phase can be calculated from the braiding matrix of conformal 
blocks is not proved although some arguments in favor of this are given in 
\cite{BW} \cite{NW}. We assume it generally holds.  
In section 2.1, we already define the  exchange  operation
 on chiral vertices:
\eqbegin 
\iPhi_{kp}^{i}(z_1)\iPhi_{pl}^{j}(z_2)=\sum_{q}R_{pq}\left[\begin{array}{cc}
 i&j\\ k&l\end{array}
\right]\iPhi_{kq}^{j}(z_{1})\iPhi_{ql}^{i}(z_{2})
\label{eq:B2}.\eqend
This is our definition of statistics. This means statistics is considered as 
a base change of the space of wavefunctions. 
Among the matrices $R$, the most important one is 
\eqbegin
R_{pq}\left[\begin{array}{cc} i&i\\ i&i\end{array}
\right].
\eqend 
This matrices should be considered as the statistics for the excitations
 $\phi_i$.
However other matrices also should be taken as the definition of statistics
 since the 
fundamental object is chiral vertex as explained above.

Let us  consider the state with $2n$ identical quasiparticles. We can consider 
the exchange or braiding of $n$ pairs of quasiparticles for this state.
Generally the expression in terms of chiral vertices 
 for the creation of $2n$ quasiparticles  has the degeneracy $D$
obtained from the fusion rules.      
As the braiding is half-monodromy around the other particle, 
this operation can be thought to be $\pi$ rotation, therefore  induces 
 a representation of $so(2)$ except the overall phase.
Those $so(2)$ can be seen as subalgebra of $so(2n)$ and forms a basis of 
Cartan subalgebra of $so(2n)$.    
 Thus  $n$ $so(2)$ representations  specify  
the highest weight of a representation of  Lie algebra $so(2n)$. 
Also the overall phase determines a representation of  $U(1)$. Therefore the  
$D$ dimensional vector space of degenerate expressions  forms thus specified  
representation of $SO(2n)\times U(1)$.
We can see this phenomenon as a extension of statistics to 
a $D$ dimensional (spinor or ordinary ) representation of $SO(2n)\times U(1)$. 
Also when we consider  the state with $2n+1$ 
identical  quasiparticles, we end up with a $D$ dimensional 
(spinor or ordinary  ) representation of $SO(2n+1) \times U(1)$. 
Whether these representation are  spinor or not 
is determined by the matrix $R$.

This extention of statistics is first introduced  
in \cite{NW} for Pfaffian states by explicit calculation and proved 
for four quasihole state. We will give a general RCFT proof for 
$2n$ quasihole  state in sect.5.2 based on the argument above.

We now discuss  the relation between nonabelian statistics  and
 the partition function of the edge state of the disk. Let us consider
 the system on the disk.
At the finite temperature $1/\beta$, the system on the  edge  is
 the same as the bulk system on the torus. 
The partition function of the edge state at finite temperature 
$i\beta=\tau$ is equal to
 the sum of  Virasoro characters (except constant): 
\eqbegin 
\chi_i={\rm tr}_{[\phi_i]}(q^{L_0+\varepsilon})
\eqend 
where $q=e^{2\pi i \tau}$ and $\varepsilon=-\frac{1}{24}c$ and $\tau$ is the 
purely imaginary modular parameter.  The modular transformation $S$ is 
\eqbegin
S: \tau  \rightarrow -\frac{1}{\tau}.
\eqend 
The behavior of $\chi_i$ under $S: \tau  \rightarrow -\frac{1}{\tau}.$ 
 transforms as a  unitary representation :
\eqbegin 
         \chi_i  \rightarrow \sum_{j}S_i^{j} \chi_j .
\eqend 
There is the remarkable  relation between the matrix
 $S_i^{j}$ and the fusion rule coefficients $N_{ij}^{k}$ of RCFT. 
\eqbegin 
N_{ij}^{k}=\sum_n S_j^{n} \lamda_i^{(n)} S_n^{k \dag} 
\label{eq:Ver}
\eqend 
\eqbegin 
\lamda_i^{(n)}={S_i^{n}}/{S^{n}_{0}}
\eqend 
This remarkable fact is
 conjectured in \cite{Ver} and proved in \cite{MorSei}(In 
\cite{Ver} Verlinde states this relation as  the modular transformation 
$S$ diagonalizes the fusion rules.). 
In our context (\ref{eq:Ver}) means the $F$,  which is the 
source of nonabelian statistics 
 is determined by the modular behavior of
 partition function of the edge excitation.  
This is to say, nonabelian statistics manifest itself in the modular propety 
of partition functions of the edge states
\footnote[1]{The modular properties of the partition function of 
 the edge excitations for the annular geometry was studied in 
 the second paper of \cite{AC2} for composite fermion constuction. }.

\subsection{Operator Algebra and Vacua}
Now let us consider what is essential to  determine the 
 phase of the given system. That is {\it vacuum} (or the ground state).
What information do we need to characterize the  vacuum  of the  phase ? 
 That is the quantum numbers of its  excitations and their  interaction. 
As we are considering the systems which are characterized  by 
some RCFT,  the statistics of excitations is determined from fusing $F$
 and conformal 
weights as in (\ref{eq:BF}). Also we will see in sect.4 that 
the interaction between the quasiparticles are   
obtained from these information in many-body system interpretations.
Thus  it is clear that it is fusion rules and Virasoro primaries  (i.e.  
operator algebra) that characterize the  vacuum  of the phase.

\subsection{Low Energy Physics} 
Let us  consider the low energy physics of the given system. The excitations which 
covern the long wavelength physics are the 
excitations which have lowest energy from vacuum. 
In our description,
it means the eigenstates of $M$  (\ref{eq:mass}) whose eigenvalue is lowest. 
Now let  $\phi_{i}^{L}$ to be lightest ones among ${\phi_{i}}$. In general, 
when the system 
has a symmetry of a group G, $\phi^{L}_{i}$ forms a representation of G. 
For example when the system is spin-singlet, the system should have $SU(2)$
 symmetry. In this case  
$\phi^{L}_{i}$ can be labeled as $\phi^{a}_{i},a=\uparrow , \downarrow$.   \\

\subsection{Rational Torus and  Uniform Magnetic Field}

We consider in particular how to deal with a system in a uniform
 magnetic field in this section.
 The problem is how one can reproduce the effect of the  uniform magnetic
 field by RCFT point of view. In this section,  we show the factor called 
the neutralizing background field  used in \cite{MorRe}\cite{BW}\cite{NW} are 
deduced from the symmetry consideration and the fusion rules of rational torus.

When a charged system is coupled to 
 a background uniform magnetic field,  there should be U(1) gauge symmetry. 
When one fixes the gauge, global U(1) symmetry remains.
 In the spirit of current algebra, U(1) current algebra naturally arises.
The theory should be described by some effective theory 
 with U(1) current algebra in the low-energy region. As  we consider a
 incompressible 
system which are characterized qualitively by $M_0$, extended U(1) algebra
 should appear.

So first let us  review extended U(1) current algebra. It is described by 
the  chiral boson  
field $\varphi$ which is compactified on a circle with a rational value 
of the square of radius. They have U(1) current algebra 
as symmetry with some extended operator. 
It has $N$ primary fields $[\phi_p]$ with U(1) charge $p/\sqrt{N}$ $ p \in Z (
{\rm mod} N)$. The extended operator 
has conformal spin $N/2$ and the unit U(1) charge. This operator corresponds 
to 't hooft operator in Chern-Simons theory \cite{MorSei2}. 
This means  the extended  
corresponds to a singular gauge transformation. As we discuss with the
  gauge  fixed,
the spectrum $p$ which appears  is  not  modulo $N$. 
This redundancy is reduced  when we take account of the singular gauge
 transformation. 
In the case of odd $N$  
(this is the case for abelian FQHE) the theory is not
 well defined on arbitrary Riemann surfaces  without additional strunctures.  
So when $N$ is odd,  we implicitly assume the space is simply connected or
 additional structure exists . 
The charge current is 
\eqbegin 
J_{z}= i\sqrt{f} \del_{z} \varphi, 
\label{eqc:urrent}
\eqend
where  $ f=\frac{1}{N}$. So $\phi_p$ has the charge $p/N$. Thus 
 $\phi_p$ are the operators for fractionally charged excitations.  

 The fusion rules 
follow  from U(1) charge conservation:
\eqbegin 
\phi_p \times \phi_q = \phi_{p+q} 
\label{eq:fuseU1}
\eqend 
This fusion rules have the same form of multiplication as  U(1). 
From the explicit expression of vertex operators  by chiral boson $\varphi$,
 we see that $\sqrt{f}\varphi $ plays the role of
 the generator  of U(1) transformation.  
Let us recall U(1) Chern-Simons theory.
 The constraint for   the  gauge fixing $A_{0}=0$ was 
 the field strength $F_{z\bar{z}}$ and $F_{z\bar{z}}$  is
  the  generator of  gauge symmetry .  Magnetic field is nothing but  a background field
 of Chern-Simons gauge field strength.    
Therefore, if we compare these two theory, 
 it is  clear that we should see $\sqrt{f}\varphi$  as  $F_{z\bar{z}}$,
 which is the field strength in Chern-Simons theory. 
So the appropriate operator 
to describe the background uniform magnetic field by U(1) current algebra
 should be 
\eqbegin 
{\rm exp}\int \frac{d^2z}{2\pi i}\sqrt{f}\varphi(z)  
\label{eq:Jiba}.
\eqend
When $N$ is odd, (\ref{eq:Jiba})  in fact reproduce
 the correct lowest-Landau-level factor of Laughlin wavefunction\cite{MorRe}.
(Note that we need to do a singular gauge transformation to go back to the 
symmetric gauge.  This is also neccesary in Landau-Ginzhburg type  theory
deviation  \cite{SCZ} of Laughlin wavefunction). 

The magnetic factor (\ref{eq:Jiba}) also can be interpreted as the 
neutralizing background field. This interpretation is understood when one 
see th operator in  (\ref{eq:Jiba}) as a charged excitation. 
Quasiparticles become  quasiholes in the presence of 
 the factor (\ref{eq:Jiba}).  

Now let us consider in a  geometrical viewpoint.
 We specify U(1) current algebra 
to be  the chiral boson $\varphi$ compactified on the circle of $({\rm radius})^2=1/N$. 
Then the vertex operators $\phi_{p}$ are 
\eqbegin
\phi_p(z)=e^{ik_p\varphi(z)},
\eqend
where $k_p=p/\sqrt{N},p=0,1\cdots N$ and  
$\phi_N$ is the extended operator. These vertex operators can  be
 interpreted  
to represent the rotation of charged particles (or the flux of 
magnetically charged quasiparticle.)
 in a uniform magnetic field. In other word 
by realizing the U(1) symmetry by U(1) current algebra, the rotation of 
charged particles are realized as  a internal structure  i.e. as a fiber 
 ( in this case, circle ). In this way,  quasiparticles have a internal
 space 
which is suppressed to be a flux in the long-wave length limit. 
And the area enclosed by the circle is $f=1/N$ times the area of the
 unit circle. The boson compactified on the unit circle is equivalent 
 two Majorana fermion i.e. Dirac fermion.  It might be possible that 
 one can take it as a usual electron (or hole ) in the integer quantum
 hall effect of lowest Landau level. 
  If it is so,  one can take $f$ as  the filling fraction itself,  
since  the ratio of the area  of circles  equals to 
 the ratio of magnetic flux .  In fact  the extended operator 
\eqbegin
\phi_N={\rm exp}i\sqrt{N}\varphi = {\rm exp}\frac{i\varphi}{\sqrt{f}}
\eqend 
is equivalent to a vertex operator of the boson compactified on
the unit circle. If $N$ is odd, the statistics 
for $\phi_N$ is fermion. Also this operator has the  unit charge. 
It is natural  that the electron ( charge  one  fermion) is
 represented by this  extended operator in U(1) current algebra
 at least when  $N$ is odd . 
Then the ratio of enclosed circles $f$ equals to the filling fraction $\nu$.
The  duality of c=1 CFT  corresponds to the duality of
 the coupling constant (or filling fraction $\nu$) around 
$\nu=\frac{1}{2}$ of abelian 
 Chern-Simons theory description of FQHE .
Among the quasiparticles, the lightest one is 
\eqbegin 
\phi_1={\rm exp}\frac{i}{\sqrt{N}}\varphi={\rm exp}i\sqrt{f}\varphi .
\eqend
This is the quasihole  operator of \cite{MorRe}.  This quasiparticle 
governs the long-wave length physics. 

This construction is readily generalized to the multiple rational torus
 compactified 
on n-dimensional lattice and reproduces the results of the hierarchical
 construction \cite{Hal} \cite{Hal2}. The physical condition must
 be imposed to ensure  the net symmetry 
of the system  is U(1) in those cases. 

We now generalize our argument  to the case
 in which RCFT has other kind of degree of freedom 
 in addition to  the extended U(1) current algebra.  We don't assume this 
additional degree of freedom is local. For example
nonlocal pairing acually occurs   in the case of Ising model .    
 We assume these additional degree of freedom to be 
 some representation ${\cal R}$ of some  group $G$.  
$G$ may be $SU(N)$ when the additional degree of freedom is isospin.   
As before U(1) part can be realized by the chiral boson  $\varphi$.
The U(1) current becomes  
\eqbegin 
 J_{z}= i({\rm dim}{\cal R})\sqrt{f} \del_{z} \varphi . 
\label{eq:Curr}
\eqend  
The field which has the unit charge have the part from the rational torus as 
\eqbegin 
{\rm  exp}\left(i\frac{\sqrt{N}\varphi}{{\rm dim}{\cal R}}\right).
\label{eq:unit}
\eqend  
The effect of magnetic field  now  amount to be 
\eqbegin 
{\rm exp}\int \frac{d^2z}{2\pi i} {\rm dim}{\cal R}\sqrt{f}\varphi(z) .  
\label{eq:Jiba2}
\eqend
This operator create the dim$R$ times larger  magnetic field
 than (\ref{eq:Jiba}).
So in this case the filling fraction  $\nu$  is not $f=1/N$. 
It now becomes 
\eqbegin 
\nu=({\rm dim}{\cal R})^2 f .
\label{eq:Nu}
\eqend 
This construction is useful  when 
 one sees RCFT as a degree of freedom of  a system in a background 
uniform magnetic field. In that case we can couple U(1) extended algebra 
to the given RCFT to achieve some  physical conditions.  
This formulation  not only describes the long-wave length physics,
but also microscopic dynamics i.e. the rotation of charged particle. 
The reason  for this feature is that 
 the symmetry  restricts the internal structure.
 This is a common feature of the worlds  described by CFT.

\section{ Quantum Mechanics of  Chiral Vertices and 
Conformal Blocks as 2+1 Dimensional Wavefunctions }
\sectionnumbering
First let us consider when  a n-point function
$ \langle  \phi^{i_1}(z_1)\phi^{i_2}(z_2)\phi^{i_2}(z_2)\cdots
\phi^{i_n}(z_n) \rangle$ 
 on a disk  can be seen as  the   ground-state wavefunction
 of some 2+1 dimensional system .
 Without knowing 
quatum-mechanical hamiltonian, it is an amplitude for that quasiparticles 
$\phi_i$ to be at $z_1\cdots z_n$. 
However some conditions must be imposed to 
be able to see this amplitude as the   ground state wavefunction of a possibly 
real physical systems.  To apply to realistic system, 
the primary field  $\phi$ which appears in the 
ground state  should be boson or fermion with an integer charge.
This gives constraints to $\phi$ which appears in a  ground-state wavefunction.  
Also this ensures us the single-valuedness of 
$ \langle  \phi(z_1)\phi(z_2)\phi(z_2)\cdots\phi(z_n) \rangle$.    
Generally  a primary field in a  given RCFT 
 can be made to have these properties  by combining them  with a vertex
 operator
 of a suitable rational torus if the primary field is a 
so-called 'simple current' \cite{BW}. 
Simple current  is a primary field which has 
a  unique fusion rule with any other primary fields and 
consequently has an abelian braiding matrix.      
When the system is under a background unform magnetic field, 
introducing the factor (\ref{eq:Jiba2}) in the previous section 
reproduces the effect of magnetic field.        

We'd like to give a simple hamiltonian on the space of chiral vertices 
 which have thus obtained  wavefunction as the  ground state.
Let $\Omega $ be the space formed by all the canonical expressions of chiral 
vertices \[\iPhi^{i_1}(z_1)\iPhi^{i_2}(z_2)\cdots \iPhi\cdots\iPhi^{i_n}(z_n)
\] where  we have omitted the internal indices  and 
boundary states. The 
omitted internal indices are sewed to  some other internal indices as 
in (\ref{eq:Gmono}) .
Generally, on  $\Omega$, we can define
 {\it Quantum Mechianics for (chiral) Vertices (QMV)} by a 
  hamiltonian  for some function $V(M,\overline{M})$  
\eqbegin 
{\bf H}=\frac{1}{2m}\del_{z_{i}}\del_{\overline z_{i}}+V(M,{\overline M}),
\label{eq:GHamil}
\eqend
where $M$ and ${\overline M}$ were defined in (\ref{eq:Mass}). 
This hamiltonian   have some useful propeties . First of all  it does not 
depend on the space. Rather it acts on the  vector  bundle $\cal V$
 over the  moduli space $\overline{\cal M}$ and its action is
 compatible with the factorization of $\cal V$.  
Second  useful property  of {\it QMV}  is that 
nonabelian statistics is manifest. It is not explicit in the
 ordinary expression of conformal blocks. 
$V(M,\overline{M})$ is separated  if the operator algebra can be  divided 
into several disjoint subalgebra. For example, when we couple rational 
torus part, $V=V_{I}+V_{torus}$ where $V_{I}$ act on internal degree of
 freedom and $V_{torus}$ act on rational torus part. 
It is also useful to define the  $\Omega_g $ which is 
 the space formed by all $ 
\iPhi^{i_1}(z_1)\iPhi^{i_2}(z_2) \cdots\iPhi\cdots\iPhi^{i_n}(z_n) $
 which express analytic blocks on genus $g$ Riemann surfaces. {\bf H} can 
be restricted on $\Omega_g$. 
When the background magnetic field, the elements in $\Omega (=\bigoplus_g 
\Omega_g) $  or 
$\Omega_g$ should be multiplied by a magnetic factor (\ref{eq:Jiba2}) and 
$\del_{z}$ in  {\bf H} should be replaced by $D_{z}$. Only the chiral half 
of operator algebra remains in the symmetic gauge.   

To see how  (\ref{eq:GHamil}) works,
 let us consider the system on a disk 
described by rational torus with $N=q$ ($N$ is defined as in sect.3.5).  
Its primary fields are labeled by $p \in Z_q$ and
 we label the extened operator as $q$. The extended operator has 
the unit charge and the statistics of fermion or boson as $q$ is odd or even. 
Then we can form  the wavefunction which can be seen as a wavefunction of 
charged boson or electron system:
\eqbegin 
\langle e^{i\sqrt{q}\varphi(z_1)}e^{i\sqrt{q}\varphi(z_2)}
\cdots e^{i\sqrt{q}\varphi(z_n)} \rangle. 
\label{eq:Laugh}
\eqend  
This is known to become Laughlin wavefunction
 for $\nu=\frac{1}{q}$ when $q$ is odd with  
the magnetic factor (\ref{eq:Jiba}) \cite{MorRe}. 
(\ref{eq:Laugh}) can be 
written in the canonical expression  of chiral vertices as   
\eqbegin 
\Psi=\iPhi^{q}_{-nq,-(n-1)q}(z_1)\cdots\iPhi^{q}_{-2q,-q}(z_{n-1})
\iPhi^{q}_{-q,0}(z_n).
\eqend 
The eigenvalue  of $M$ for  this wavefucntion is ${\cal E}_{\Psi}
=\frac{n(1-n)q}{2}$. $-E_{\Psi}$ equals to the total multiplicity of zero 
of Laughlin wavefunction. 
The creation of $\phi_p$ are given by multiplying ,
  $\iPhi^{p}_{-p-nq,-nq}$ from left(this corresponds to create $\langle 
\phi_p\phi_e \cdots\phi_e\rangle$).   By the definition of $M$,
\eqbegin 
M\iPhi^{p}_{-p-nq,-nq}=-np\iPhi^{p}_{-p-nq,-nq}
\eqend 
Then, if we take $V$ to be
\eqbegin 
 V(M)={\cal E}_{\Psi}-M,
\label{eq:VV}
\eqend 
(\ref{eq:Laugh}) is the exact zero-energy
state of ${\bf H }$. (\ref{eq:VV}) counts
 the increase of the total multiplicity of zeros of 
wavefunctions.  The gaps of the excitations above this zero-energy state are  
\eqbegin 
V(M)\iPhi^{p}_{-p-nq,-nq}(z_j)\Psi=
np\iPhi^{p}_{-p-nq,-nq}\Psi 
\eqend 
This value does not depend on where we put $\iPhi^{p}$ in the conformal block.
For $\iPhi^{p}$ which can be multiplied at some  boundary state, 
the excited  energy can be obtained by acting $V(M)$ on $\iPhi^{p}$ alone.    
Now it is clear that $\phi_1$ is the lightest quasiparticle in 
this system.  Also, one can take similar potentials such as following one:  
\eqbegin 
 V(M)=({\cal E}_{\Psi}-M)^2.
\label{eq:VV2}
\eqend 
Generally, the operator $(-M)$ gives the total multiplicities of zero 
 or total angular momentum when it acts on a canonical expression on disk.   
   
Next,let us consider {\it QMV} in more generalities.  
Let $\Sigma$ be a genus $g$ 
Riemann surface with $m$ boundaries. General amplitude on $\Sigma$ can be 
written in terms of the element in $\Omega_g$: 
\eqbegin 
\Psi=\iPhi_{b_1 a_1}^{i_1}\cdots
\overline{\iLamda}_{d_{3j-2} d_{3j}}^{d_{3j-1}}\cdots  
\iLamda_{c_{3k-2} c_{3k}}^{c_{3k-1}}\cdots
\iPhi_{a_{2l-2} a_{2l-1}}^{i_{i_l}} \cdots 
\iPhi_{a_{2n-2} 1}^{i_{n}}. 
\eqend
Let us consider  the following potentials :    
\eqbegin 
M_a=\frac{1}{2}(M+\overline{M}) ,\\ 
M_b=\frac{1}{2}(\overline{M}-M) .
\label{eq:V_2}
\eqend
From (\ref{eq:Cal1})(\ref{eq:Cal2}),  
\eqbegin 
M_a\Psi=\left(\sum_{k} \iDelta_{i_k} \right)\Psi ,\\ 
M_b\Psi=\left(\sum_{m}\iDelta_{b_m}\right)\Psi.
\eqend
So $M_a$ only sees the excitations at the bulk. This operator plays the role 
of mass operator in general {\it QMV}.
It reproduce the description of 'mass'  in sect.3.1 at quantum mechanical level 
  (in the case of closed Riemann surface, $M$ or  $\overline{M}$ is enough).
On the other hand, $M_b$ only sees the states at the boundaries. The boundary 
states are determined by the fusion rules in the canonical 
 expression of the amplitude.
 Every potentials $V(M,\overline{M})$ can be  rewritten in terms of 
$M_a$, $M_b$.  This means one can always separate these two contributions.
Also we can see how the bulk states and  the boundary states  correlates 
each other in the expansion of $V(M,{\overline M})$ in terms of $M_a$ and $M_b$.

In {\it QMV} on a general 
Riemann surface $\Sigma$,  fusion rules are considered
 to be a kind of conservation laws. This analogy 
arises  since,  in the canonical expression of wavefunctions through 
chiral vertices, the  expression depends on 'flow' or
 Morse function on $\Sigma$ and 
 the fusion rules controls the joint and split of the  'flow' on $\Sigma$.   
In ordinary mechanics,  the conservation laws 
are determined by the symmetry group it has. 
The conserved currents form a representation of its Lie algebra. 
As explained in  \cite{MorSei}, RCFT can be seen as a generalizaion of 
group theory.  {\it QMV} has the generalization of group theory as 
undelying conservation law. Also, as in sect.3.2, fusion rules and 
mass (conformal weight), i.e. operator algebra determines the statistics 
of quasiparticle of the system.   

Degeneracy of a given wavefunction is  computed from the expression 
through chiral vertices by the fusion rules.
For example, the degeneracy of the wavefunction for Laughlin wavefunction 
for filling factor $\nu=1/q$ is shown to be $q^g$ on genus g Riemann surface.  

\section{Some examples }
\sectionnumbering

We would like to  take some specific examples in this section.  For application 
to real physics, we only consider the system in a background uniform magnetic field.
We will consider $SU(N)_k$ WZW model, and Ising model. $SU(2)_1$ model 
gives the Halperin state of FQHE \cite{Halp} and Ising model  
gives so-called Pfaffian state \cite{MorRe}. 
\subsection{$SU(N)_k$}  
$SU(N)_k$ WZW models
 with rational torus were  already  discussed as 2+1 dimensional  many-body 
system in \cite{BW}. These model have excitations with nonabelian statistics for 
$k > 1$. However we mainly consider $SU(N)_1$ WZW model in this section.  
This model gives us a generalization of Halperin state. The relation between 
composite fermion and  $SU(N)_1$ model is discussed  in \cite{AC2}. 
Our discussion is simple applications of discussions  in sect.3.5 and sect.4 .  

First  let us recall $SU(2)_k$ WZW models.  $SU(2)_k$ has k+1  
representations $[\phi_l]$,namely the ones with $SU(2)$ isospin 
$\frac{1}{2}l\le \frac{1}{2}k$. The conformal weights for $\phi_l$ are 
\eqbegin 
h_l=\frac{l(l+2)}{4(k+2)}.
\eqend 
 The fusion rules are 
\eqbegin 
\phi_l \times \phi_{\acute l} = \sum_{j=|l-{\acute l}|}^{{\rm min}(l+{\acute l},2k-l-{\acute l})} \phi_j ,
\eqend   
where $j-|l-{\acute l}|$ is an even integer. From this fusion rules, when $k > 1$,
 we have excitations with nonabelian statistics  
when this RCFT becomes the  internal degree of freedom of 2+1 dimensional system 
by coupling  a suitable rational torus.  
The modular behavior of the 
characters $\chi_l$ are 
\eqbegin 
S_{ln}=\left(\frac{2}{k+2}\right)^{1/2}{\rm  sin}\frac{(l+1)(n+1)}{k+2}\pi 
\eqend 
The  matrices $F$ and $R$ are known for these models.  See,  for example 
\cite{Schro} for the formulas of these matrices.    

Let us  now consider  $SU(2)_1$. $SU(2)_1$ WZW model has
 two primary fields $(k=1)$
\eqbegin 
V^{\downarrow}(z_{\downarrow}),      V^{\uparrow}(z_{\uparrow})=J^{+}_{0}V^{\downarrow}
(z_{\uparrow}),
\eqend 
where $J^{+}$ are the creation operator in the standard basis of $su(2)$.
The conformal weights for these vertex operators are respectively  
\eqbegin 
h_l=\frac{l(l+2)}{4(k+2)}.
\eqend 
The fact that $SU(2)$ freedom is nothing but spin degree of freedom
 suggests its relevance to electronic system. 
When the system is in a background uniform magnetic field, rational torus 
with $N=q$ ($N$ defined in sect.3.5 ) appeared as explained in sect.3.5.
From (\ref{eq:unit}),  the primary fields with the unit charge are 
\eqbegin 
V^{\downarrow(\uparrow)}(z)e^{i\frac{\sqrt{q}}{2}\varphi(z)}.
\eqend 
They are in fact
 fermion when $q =4m+2$  when  $m$ is even and boson when $m$ is odd. 
Thus this field have the  same quantum numbers  with  electron when $q =4m+2$ with 
even $m$.
We can construct the $SU(2)$ singlet ground state from these fields; 

$ \langle V^{\uparrow}({z_1}^{\uparrow})e^{i\frac{\sqrt{q}}{2}\varphi({z_1}^{\uparrow})}
 \cdots
V^{\uparrow}({z_n}^{\uparrow})e^{i\frac{\sqrt{q}}{2}\varphi({z_n}^{\uparrow})}
V^{\downarrow}(z_1^{\downarrow})e^{i\frac{\sqrt{q}}{2}\varphi(z_1)}\cdots \cdots $
\eqbegin 
\cdots V^{\downarrow}(z_n^{\downarrow})e^{i\frac{\sqrt{q}}{2}\varphi(z_n^{\downarrow})}
{\rm exp}\int \frac{d^2z}{2\pi i} \frac{2\varphi(z)}{\sqrt{q}} \rangle 
\label{Halperin}
\eqend 
where the factor 2 in the magnetic factor comes from the internal 
degree of freedom as in (\ref{eq:Jiba2}). This 
state is manifestly $SU(2)$ invariant.  Actually this is  
 the spin-singlet state, so-called Halperin state \cite{Halp} \cite{MorRe}. 

$
\prod_{i<j}(z_i^{\uparrow}-z_j^{\uparrow})^{\frac{q}{4}}\prod_{i<j}(z_i^{\downarrow}-z_j^{\downarrow})^{\frac{q}{4}} $
\eqbegin
\times\prod_{i<j}(z_i^{\uparrow}-z_j^{\downarrow})^{\frac{(q-2)}{4}}{\rm exp}\left[ -\frac{1}{4}\sum(|z^{\uparrow}_i|^{2}+|z^{\downarrow}_i|^{2})\right]
\eqend
Its filling fraction is $\nu=\frac{4}{q}$ from (\ref{eq:Nu}).
 The lightest excitation of above the state (\ref{Halperin}) is given by  
\eqbegin 
V^{\downarrow,\uparrow}(z)e^{i\frac{\varphi(z)}{\sqrt{q}}}.
\eqend 
These excitations governs the low-energy physics of the system.

Next let us consider the a system with internal $SU(N)$ symmetry.
The excitations are in the $N$-dimensional 
representation of $SU(N)$.  The system are  realized  as $SU(N)_1$ WZW model. 
The conformal weight for the primary fields is 
\eqbegin 
\frac{N^{2}-1}{2N(N+k)}.
\eqend   
where $k=1$.
In this case, by coupling  the rational torus with 
\eqbegin 
q=N^2m+\frac{N(N^2-1)}{(N+k)} ,
\eqend 
the system has a boson or fermion with the unit charge as  
$m$ is even or odd respectively. By using antisymmetric tensor, 
we can again form the $SU(N)$ invariant  ground state wavefunction.
From (\ref{eq:Jiba2}), the magnetic factor is 
\eqbegin 
{\rm exp}\int \frac{d^2z}{2\pi i} N\frac{\varphi(z)}{\sqrt{q}} .  
\eqend
The state has filling fraction 
\eqbegin  
\nu=  \frac{N}{Nm+\frac{N^2-1}{(N+k)}} 
\eqend 
from (\ref{eq:Nu}).

\subsection{Ising Model}
We'd like to cast Ising model in the present framework. 
Ising model has three primary fields  $1,\psi,\sigma$. Here $\psi$ is a
 Majorana fermion and $\sigma$ is the spin field. Their conformal weights are 
respectively $\iDelta_\psi =\frac{1}{2},\iDelta_{\sigma}=\frac{1}{16} $   .
The fusion rules of Ising model are 
\eqbegin \psi\times\psi=1 ,      
  \psi\times \sigma =\sigma  ,   
  \sigma \times \sigma =1 +\psi
\label{eq:IsFus}
\eqend 
Under  a suitable  normalization, we get the fusion matrices $F$ as follows:
\[F\left[\begin{array}{cc} \psi &\psi \\ \psi & \psi \end{array} \right]=1,
F\left[\begin{array}{cc} \sigma &\psi \\ \psi & \sigma \end{array} \right]=-1,
\]
\eqbegin
F\left[\begin{array}{cc} \sigma &\sigma \\ \sigma & \sigma \end{array} \right]=\frac{1}{\sqrt{2}}\left(\begin{array}{cc} 1 & 1 \\ 1 & -1 \end{array} \right) 
\eqend
From these formulae we get the exchange matrix $R$,
\eqbegin 
R\left[\begin{array}{cc} \sigma &\sigma \\ \sigma & \sigma \end{array} \right]
=\frac{e^{i\pi/8}}{\sqrt{2}}\left(\begin{array}{cc} 1 & -i \\ -i & 1 \end{array}
 \right)
\label{eq:IsEx}
\eqend  
Let us make a boson or fermion with the unit charge 
from $\psi$ by coupling a rational torus with suitable $N=q$. 
From the fusion rules (\ref{eq:IsFus}), $\psi$ can appear only when they are paired. Therefore 
internal degree of freedom is $Z_2$, and a pair of $\psi$ 
forms a representation with dim${\cal R}=2$. The filling fraction is $\nu=\frac{4}{q}$.
 From (\ref{eq:unit}) we 
can form a boson or fermion with the unit charge as   
\eqbegin 
\psi e^{i\frac{\sqrt{q}}{2}\varphi } 
\eqend
It is a charge one fermion when $q=8k$ and a charge one boson 
 when $q=8k+4$. From this field  we can form a ground-state wavefunction 
for a certain 2+1 dimensional quantum mechanics as explained in sect.4 :
\eqbegin 
\langle \psi(z_1)e^{\frac{i\sqrt{q}}{2}\varphi(z_1)} \cdots 
\psi(z_m)e^{\frac{i\sqrt{q}}{2}\varphi(z_m)}  {\rm exp}\int \frac{d^2z}{2\pi i}
2\frac{\varphi(z)}{\sqrt{q}}\rangle
\eqend
This wavefunction is  Pfaffian state first derived in \cite{MorRe} 
when $q=8k$. The name ' Pfaffian ' comes
 from the fact that this  wavefunction equals to 
\eqbegin 
{\rm Pfaff}(\frac{1}{z_i-z_j})\prod_{i<j}(z_i-z_j)^{\frac{q}{4}}{\rm exp}\left[ -\frac{1}{4}
\sum_i|z_i|^2\right] .
\eqend 
The lightest excitation  in this system is the quasihole  
\eqbegin 
\sigma {\rm e^{i\frac{\varphi}{\sqrt{q}}}}.
\eqend 
However this field  cannot appear alone on the Pfaffian state .This is clear
 from the fusion rules.
 We had better use the chiral vertices expression for further consideration.
Pfaffian  state can be written in terms of chiral vertices as 
\eqbegin 
\iPhi_{1\psi}^{\psi}(z_1)\iPhi_{\psi 1}^{\psi}(z_2)\cdots\iPhi_{1\psi}^{\psi}
(z_{n-1})
\iPhi_{\psi 1}^{\psi}(z_{n})
\label{eq:Pfaff}
\eqend
where the rational torus  part are omitted. 
The potential in
 (\ref{eq:GHamil}) which have (\ref{eq:Pfaff}) as its exact ground state 
is, for example, 
\eqbegin 
V(M,\overline{M}) = {\cal E_{\rm Pfaff}}-M ,
\eqend  
where ${\cal E_{\rm Pfaff}}$ is the eigenvalue of $M$ for Pfaffian state. This 
potential counts the increase of the total multiplicities of zeros of the 
wavefunctions.
There are two possible expressions composed from 
 two chiral vertices to insert into (\ref{eq:Pfaff})
 which include  the spin field $\sigma$. They create 2 quasihole state.
They are    
\eqbegin 
\Psi=\iPhi_{1\sigma}^{\sigma}\iPhi_{\sigma 1}^{\sigma}\\\Psi'=\iPhi_{\psi
 \sigma}^{\sigma}\iPhi_{\sigma \psi }^{\sigma}.
\eqend 
From (\ref{eq:V_2}), these are the lightest excitation of the 
system. Actually these two expressions create the same state up to phase. 
This situation is same for general $2n$ case, so we may only consider 
the expressions of type $\iPhi_{1\sigma }^{\sigma}\cdots\iPhi_{\sigma 1}^{\sigma}$.
For 4 quasihole state, the expressions are :
\eqbegin 
\Xi_1=\iPhi_{1\sigma}^{\sigma}\iPhi_{\sigma 1}^{\sigma}\iPhi_{1\sigma}^{\sigma}
\iPhi_{\sigma 1}^{\sigma} \\ \Xi_2= \iPhi_{1\sigma}^{\sigma}\iPhi_{\sigma \psi}^{\sigma}\iPhi_{\psi \sigma}^{\sigma}\iPhi_{\sigma 1}^{\sigma}
\eqend
This degeneracy 2 can easily expected  from the fusion rules (\ref{eq:IsFus}).
Obviously when we exchange two chiral vertices in the middle, non-abelian
 statistics appears. $\Xi_1$ and $\Xi_2$ are transformed irreducibly
 in the exchange. Thus this system has excitations with nonabelian statistics. 
From the general discussion in sect.3.2 for multiple quasiparticle state, 
this statistics can be extended to a representation of $SO(4)$.
From (\ref{eq:IsEx})  we see that the representation formed 
by $\Xi_1$ and $\Xi_2$  is 
2 dimensional  spinor representation of $SO(4)\times U(1)$.  

Next let us consider $2n$ quasihole state. As in $n=2$ case, the 
degeneracy of  the canonical expression for $2n$ quasihole state is 
 $2^{n-1}$ from the fusion rules (\ref{eq:IsFus}). 
Again from the general discussion of sect.3.2, we see that
 the vector space of these degenerated states forms 
 $2^{n-1}$ dimensional representation of $SO(2n)\times U(1)$.  
and it  is  spinor from   (\ref{eq:IsEx}). 

This result 
is proved in \cite{NW} for the four quasihole state by an explicit 
calculation of conformal block,  with the  indication for
 $2n$ quasihole state. Our discussion  doesn't depend on the explicit form 
of conformal blocks. This is natural because statistics is determined 
only from the operator algebra of RCFT. 

Among $2n$ quasihole states, 
$8$ quasihole state ($n=4$) forms 8 dimensional spinor representation of $SO(8)$.
It is interesting that this representation coincides with 
 the   space-time interpretation  for the ground state  of Ramond sector 
 of superstring. Also the triality relation  of $SO(8)$ on this state
 is interesting. These issues will  be explored elsewhere.

\end{document}